# Uncovering the role of ionic doping in hydroxyapatite: The building blocks of tooth enamel and bones

Mahdi Tavakol[1, 2], Jinke Chang[1], Cyril Besnard[1], Gabriel Landini[2], Richard M. Shelton[2], Jin-Chong Tan[1*], and Alexander M. Korsunsky[1,3**]

[1] Multi-Beam Laboratory for Engineering Microscopy (MBLEM) and Multifunctional Materials & Composites (MMC) Laboratories, Department of Engineering Science, University of Oxford OX1 3PJ, UK

[2] School of Dentistry, University of Birmingham, 5 Mill Pool Way, Edgbaston, Birmingham B5 7EG, UK

[3] Professor and Fellow Emeritus, Trinity College, Oxford OX1 3BH, UK

Corresponding authors:
[**]alexander.korsunsky@eng.ox.ac.uk
[*]jin-chong.tan@eng.ox.ac.uk

## Abstract

Hydroxyapatite (HAp) is the primary mineral component of various mineralized tissues in the human body, including bone and teeth, where it performs critical roles of structural support and load transmission. In the context of dental health, the two most crucial properties of HAp are mechanical stability, which ensures resistance to forces, and chemical stability, which preserves surface integrity in acidic environments. During early stages of human evolution, e.g. when teeth were used to crush uncooked food, mechanical stability was of paramount importance. However, with changes in diet and lifestyle, the principal origins of tooth damage and loss shifted towards bacterially mediated chemical attack, known as tooth decay, or caries. To enhance the chemical stability, ion doping has emerged as a particularly significant approach, and it lies at the focus of the present study.

A Molecular Dynamics (MD) framework was developed to investigate the effects of ion doping on the chemical and mechanical stability of HAp and to identify optimal doping candidates. The framework combines conventional MD with Steered Molecular Dynamics (SMD), Thermodynamic Integration (TI) and uniaxial compression test simulations to provide comprehensive insights into the doping process. The findings revealed surface atoms as the most viable candidates for doping, as demonstrated by SMD and conventional MD simulations.

Notably, TI calculations have identified magnesium ions as a better candidate among the ions considered here for enhancing the chemical stability of HAp. The results presented in this study offer valuable guidelines for synthesizing HAp-based substituent materials with properties tailored to meet the demands of modern dental applications such as implant coatings, enamel remineralization agents and restorative materials.

## Introduction

Hydroxyapatite (HAp) with the chemical formula of $Ca_{10}(PO_4)_6(OH)_2$ for its stoichiometric pure form is the main mineral of tooth enamel and bone [1]. Among various properties of HAp, the mechanical and chemical stability are central to the functions of the organs and tissues made of it. Mechanical stability pertains to the material's ability to withstand forces without fracturing while chemical stability refers to the material's resistance to dissolution and degradation. The relative importance of each one of these properties in each organ and tissue is determined by the characteristic function of that organ. In teeth, for example, the mechanical properties mattered the most at the early stages of evolution so that the tooth could resist the high forces occurred in eating hard foods. Nowadays, however, given the diet is softer and rich in carbohydrates, the chemical stability of HAp has become more important, where HAp must resist acidic conditions from dietary sources and microbial fermentation of those dietary carbohydrates [2]. Ensuring the chemical stability of HAp is essential for preventing caries, along with the consideration of the possibility of remineralizing enamel lesions, and promoting overall dental health [3, 4].

Despite the undeniable importance of mechanical properties such as compressive strength and high strength, HAp is found to be somewhat deficient in maintaining the chemical balance necessary for tooth remineralization [5]. Therefore, prioritizing the chemical stability of HAp is crucial for effective dental health maintenance, treatment and the long-term preservation of tooth structure. To strike a balance between chemical and mechanical stability, the chemical content of HAp can be tailored through ionic doping that forms the topic of the present study. Ions can be incorporated into the tooth structure to enhance the chemical stability of HAp, improving its resistance to demineralization [6].

HAp doping with various ions has been previously investigated utilizing mostly experimental techniques, with studies considering specific dopants due to their unique properties. For instance, the antimicrobial properties of zinc (Zn) and silver (Ag) motivated several studies to investigate either Zn- or Ag-doped HAps [7-11]. In the case of Ag doping, better antibacterial protection against gram-positive and gram-negative bacteria was observed [9, 10]. Various synthetic methods such as co-precipitation and pulsed laser deposition have been considered and optimized to achieve a uniform distribution of silver ions that is important for the structural integrity of HAp [10, 11].

Magnesium, fluoride and carbonate have also been extensively researched in the context of ionic doping of HAp [12-22]. The incorporation of $Mg^{2+}$ at the level of 1 mol% has been shown to enhance bioactivity, particularly by supporting the proliferation and differentiation of osteoblastic cells — key processes involved in the early stages of bone regeneration [12, 13]. Introducing fluoride ions into the HAp structure leads to the formation of fluorapatite, which has improved stability and lower solubility compared with pure HAp. This modification has been reported to improve bonding strength and promote the deposition of bone-like apatite, both of which are favorable factors that may contribute to improved osseointegration in dental and orthopedic applications [15, 16]. Additionally, fluoride-doped HAp has been shown to promote dental remineralization, effectively occluding dentinal tubules and reducing sensitivity [17]. Incorporating $CO_3^{2-}$ into the HAp lattice can enhance its solubility and bioactivity, mimicking the natural mineral composition of bone [20, 23-27]. Studies have shown that carbonate-doped HAp exhibits improved osteoconductivity and promotes mineralization processes compared to undoped HAp [23]. The presence of carbonate ions can also influence the crystallinity, morphology of HAp and mechanical properties [25-27].

In summary, in the context of modern healthcare the chemical stability of HAp has become relatively more important for tooth enamel. Despite numerous studies investigating the effects of ionic doping on the HAp mineral, many aspects remain unclear, e.g., the impact of various ions on different facets of HAp crystals across a range of pH values and ionic concentrations. The present study addresses these gaps by investigating the microscopic details of ion doping that remained elusive and are lacking in literature to date through a combination of both the computational and experimental approaches. This study reveals a more pronounced effect of $Mg^{2+}$ compared to the fluoride and carbonate ions.

# Results

The results section is organized in several subsections. First, the results of the conventional MD simulations are summarized to determine the preferred location of the dopants inside HAp. The results of SMD simulations are presented in which the kinetics of the ion doping are revealed. Finally, the last two subsections contain information on the effect of dopant on the thermodynamics and mechanical stability.

**Conventional MD simulations**

The conventional MD simulations for 001 and 010 facets of HAp at pH values of 5 and 7 solvated in a $Mg_3(PO_4)_2$ solution were carried out to investigate the possibility of the ion doping in the HAp. These two pH values were simulated because at pH 5, the HAp is composed of dihydrogen phosphate while there is a combination of both monohydrogen and dihydrogen phosphates at pH 7. The adsorption of the $Mg^{2+}$ ion on the surface of HAp is observed through ionic density profiles and simulation snapshots while there is no sign of ion doping inside the HAp surface. Even increasing the salt concentration to the values as high as 2M does not lead to ionic doping inside HAp (Figure S1). Three possible reasons for this observation are either (a) the ions doped on the surface are more stable than those in the interior, (b) the limited timespan of the conventional MD simulation did not reveal doping inside HAp, or (c) the ionic doping inside HAp is not possible without ionic demineralization. To test the first hypothesis, conventional MD simulation with ions substituted inside HAp were performed to see if the ions inside the HAp move to the surface. These simulations did not show such a possibility which highlights the importance of the second and third hypothesis. Regarding them, the first step of the ion substitution inside HAp requires the creation of a void inside it. The possibility of ion substitution inside HAp in the simulation time or over longer timescales depends on the availability of the activation energy of such a void which is studied in the next subsection.

**SMD simulation**

SMD simulations were carried out for 001 and 010 facets of HAp at pH values of 5 and 7 (Figure 1). The free energy profiles illustrate that the pulling distance for each case is large enough for the pulled ion to evade the interaction sphere of the HAp (Figure 1a). The minimum free energy values show that the free energy of the surface ions is at least an order of magnitude lower than the rest of the ions making them the suitable candidate for the dissolution from the HAp creating the vacancies required for the ion doping (Figure 1b). Even though this observation does not exclude the possibility of the ion doping in the HAp interior, the higher free energy of the ions in deeper layers of HAp suggests that the surface ions are only viable candidates for ion doping. These energies also hint at the possibility of HAp dissociation through a row-by-row process rather than the formation of dislocations perpendicular to the HAp surface.

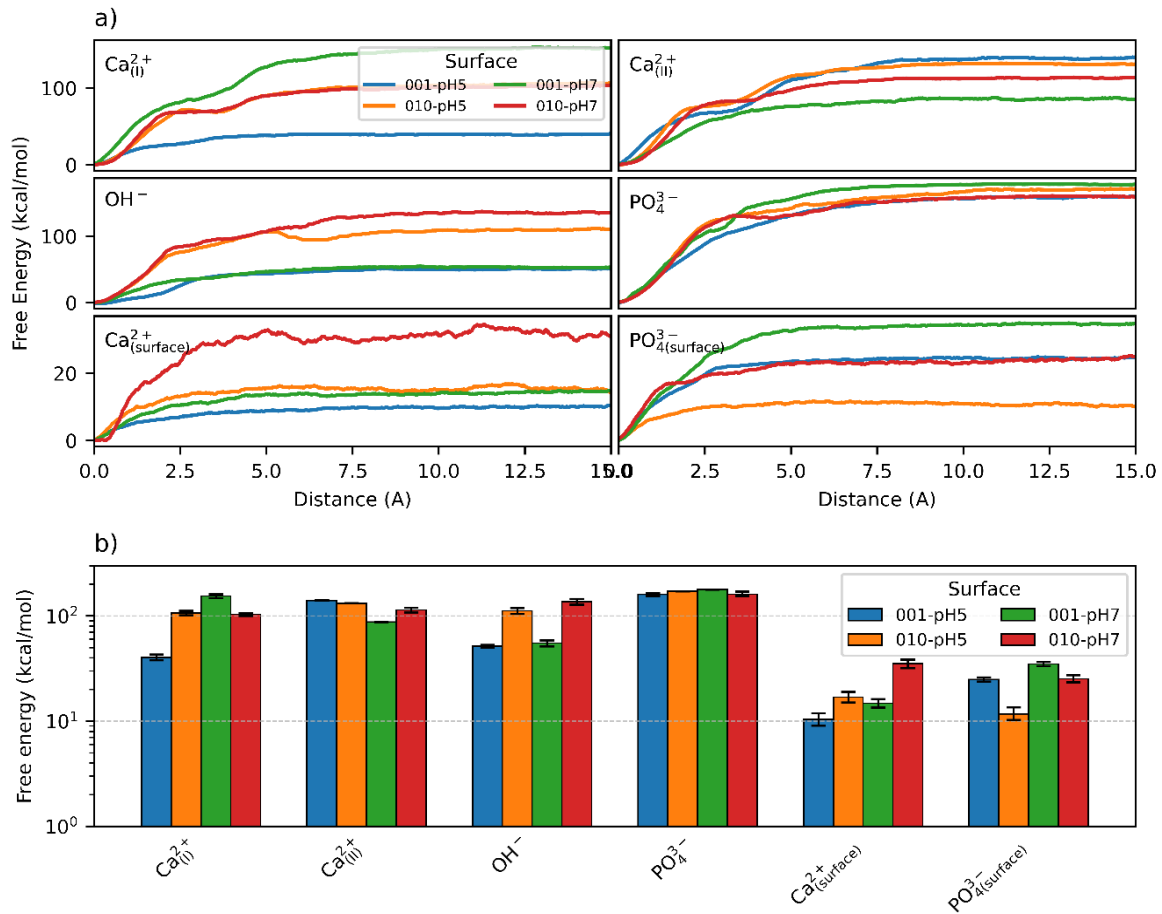

Figure 1 – SMD simulation results showing the kinetics of the ion doping. a) The SMD free energy profiles obtained from the average work for the forward and reverse paths. The distance is measured from the initial position of the atom being pulled away towards the solution. b) The value of the minimum free energy profiles which provides information about the plausible ion dissociation pathways from the HAp surface.

### Chemical Stability

The chemical stability values shed light on the thermodynamics of ion doping in HAp. The chemical stability values as a result of different doping extent of (in atomic percent) 10%, 20% and 50% for three ion substitution of $Ca^{2+}$ to $Mg^{2+}$, $OH^-$ to $F^-$ and $PO_4^{3-}$ to $CO_3^{2-}$ are depicted in Figure 2a. The highest effect is seen for the substitution of $Ca^{2+}$ with $Mg^{2+}$ which increases with the percentage from 10% to 30%, observing a change in the chemical stability from the value of -0.37±0.06 kcal/mol to -2.14±0.27 at pH 5. For the doping of $PO_4^{3-}$ with $CO_3^{2-}$ and $OH^-$ with $F^-$ the chemical stability does not change significantly even for the highest percentage of 50%. To decipher the effect of solvation from the surface, the change in the free energy of HAp with the ionic substitution is also obtained (Figure 2b). The HAp free energy for the 50% $Ca^{2+}$ to $Mg^{2+}$ replacement at pH 5 is -10.35±1.76 kcal/mol which is higher than the value of -2.14±0.27 kcal/mol for the chemical stability. Interestingly, the HAp free energy change reaches a positive value of 1.75±1.1 kcal/mol at pH 5 for $PO_4^{3-}$ doping with $CO_3^{2-}$ despite the negligible chemical stability for this case, and finally the HAp free energy for the $OH^-$ doping with $F^-$ is zero. Thus, for the $Ca^{2+}$ doping with $Mg^{2+}$ the solvation free energy becomes less favorable decreasing the chemical stability while for the for the $PO_4^{3-}$ doping with $CO_3^{2-}$ the solvation is more favorable counteracting the unfavorable change in the free energy of the HAp, and solvation free energy does not change for the substitution of $OH^-$ with $F^-$.

Comparison of ionic doping for the 001 surface with different pH values show that considering the error bars, the pH neither affects the HAp free energy nor the chemical stability of the system. This is expected considering the fact that just the protonation of the surface phosphate ions changes as a result of pH changes, and inside HAp where ionic doping mostly occurs the structures remain the same. There seems to be a trend towards pH effects on the HAp free energy for higher percentages which might be due to the higher possibility of the ion doping occurring within the surface where the protonation change due to variations in pH occurs.

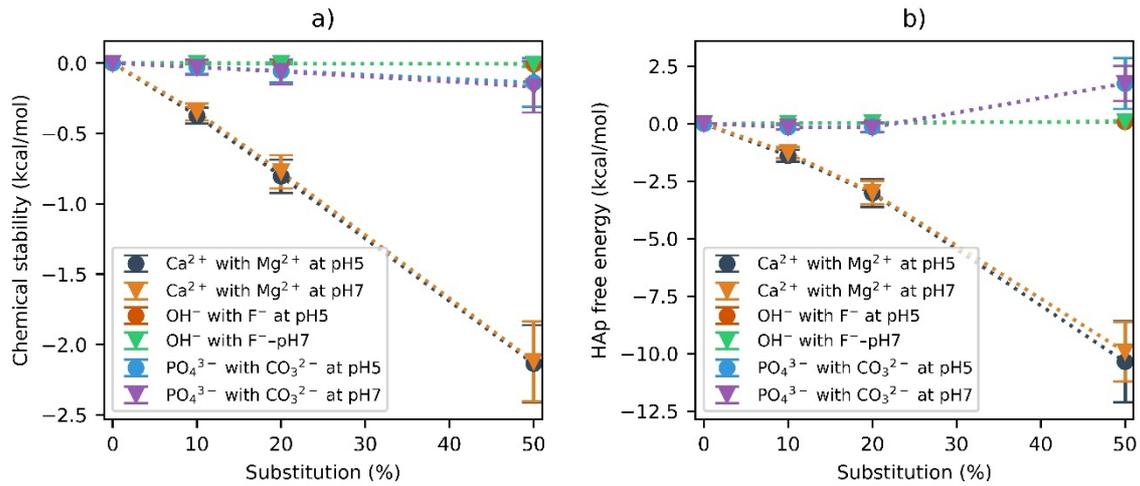

*Figure 2 – Change in the chemical stability of HAp as a result of doping of various ions with different concentrations showing various influencing factors. a) HAp free energy and b) the chemical stability of the doped HAp system as a function of different doping concentrations and ions.*

## Mechanical stability

The final aspect of the ion doping investigated in the current study is the mechanical properties defined as the resistance towards mechanical forces. The stress-strain curves for all ion doping variants and percentages considered here showed a drop in the stress after the Ultimate Compressive Strength (UCS) and a final constant stress (Figures 3a, b, c). The simulation snapshots illustrate the sudden change in the crystalline to the amorphous structure at the UCS point which may be considered the reason for the drop in the stress value. The final constant stress is the artifact of the simulation method as our simulations with solvated HAp structure did not show such a behavior.

The UCS and stiffness were taken as the measures of mechanical stability and obtained through analysis of the stress-strain curves (Figures 3d, e, f). Interestingly, the UCS and the stiffness illustrate the largest change for $Ca^{2+}$ doping with $Mg^{2+}$ which increases with the doping percentage with a decrease of the stiffness from 141.44±0.16 GPa for 5% to 91.43±1.09 GPa for 30% substitution. For the other two cases of $OH^-$ doping with $F^-$ and doping of $PO_4^{3-}$ doping with $CO_3^{2-}$ there is no noticeable change in the mechanical stability.

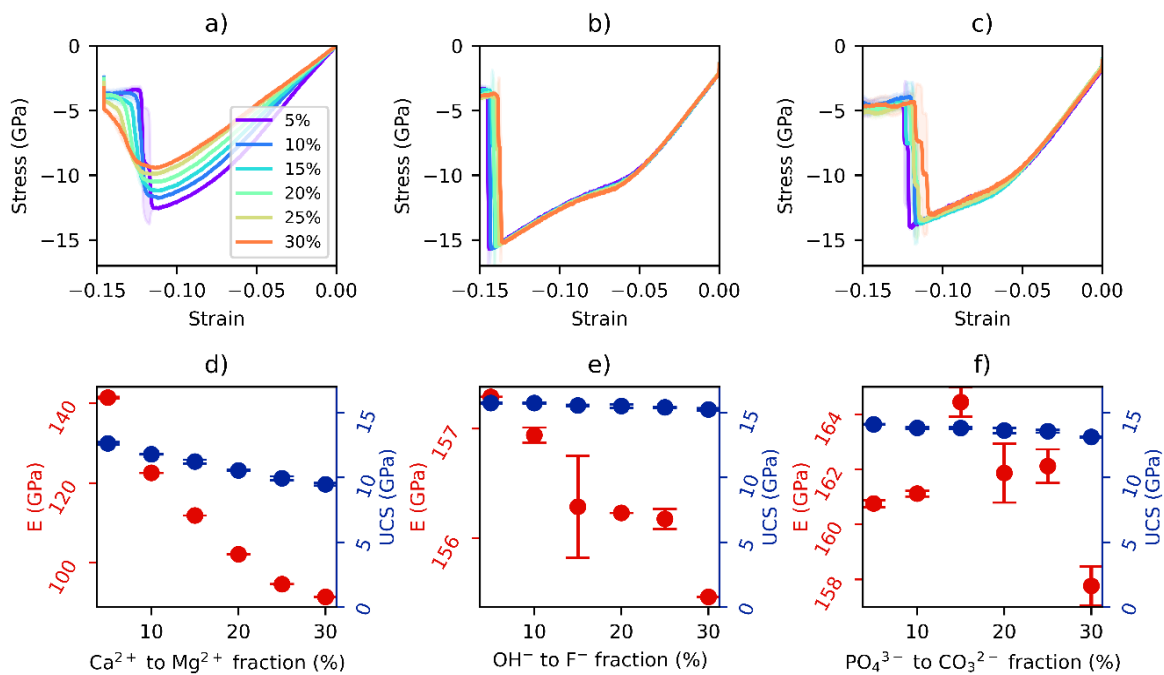

*Figure 3 – Uniaxial compression simulation results for the HAp systems with various doped elements and concentrations illustrating changes in the HAp mechanical properties. (a), (b), (c) Stress-strain plots and (d), (e), (f) UCS and stiffness values for (a), (d) $Ca^{2+}$ doping with $Mg^{2+}$, (b), (e) $OH^-$ doping with $F^-$ and (c), (f) $PO_4^{3-}$ doping with $CO_3^{2-}$ with various doping concentrations.*

## Discussion

HAp is the main mineral of bone, tooth enamel and various mineralized organs in the human body. The chemical stability is particularly important for tooth enamel compared with its mechanical stability in the oral cavity, as it is required to resist dissolution in the acidic conditions from dietary sources or microbial activities [28, 29]. Ion doping is a significant approach to enhance chemical stability which can also be used to enhance other properties too. Despite the wealth of information available in the literature on the effect of HAp doping with various ions [7-22, 30-35], it remained hitherto unclear how the chemical and mechanical stability of HAp changes with the doping percentage, whilst the molecular mechanisms of doping were insufficiently well understood. The current study presented novel insights into these aspects of ion doping. First, the location of doped ions was determined through conventional MD and Steered Molecular Dynamics (SMD), and then the chemical and mechanical stability aspects of the $Mg^{2+}$, $OH^-$ and $CO_3^{2-}$ doped HAp were systematically evaluated through Thermodynamic Integration (TI) and the uniaxial compression test simulation.

Despite various studies dedicated to HAp doping with different ions [7-22, 30-35], the unanswered question from the literature survey was where the ions end up in the HAp structure, with important implications for the molecular evolution of the doped HAp species in the biological media. This is summarized below. Neither for the $Mg^{2+}$-doped HAp nor for the pure HAp solvated in a high concentration solution of $Mg_3(PO_4)_2$ solution was there an exchange of $Mg^{2+}/Ca^{2+}$ ions observed between the interior of HAp and ions in the solution. This has two implications: either the simulation duration was insufficient, or this doping process is not possible. These hypotheses were tested, noting that the first stage of bulk HAp ion doping is the creation of vacancies within HAp to place ions there. The free energy of this process is the activation free energy of the doping process. The SMD simulations deployed calculated the activation free energy and investigated the possibility of ion doping for different atomic coordinates inside HAp. The free energy values for desorption of a $Ca^{2+}$ ion from the first HAp ionic layer from 001 and 010 facets at pH values of 5 and 7 were found to be -9.13, -12.44, -12.55 and -36.9 kcal/mol. Equation 8 relates the time required to pass the barrier ($t$) with $A$ which is in the order of $10^{13}$ s$^{-1}$ for many reactions, and $E_a$, $k$ and $T$ representing the activation free energy, the Boltzmann's constant and the temperature. Accordingly, these free energy values translate to the time scales of $4.33 \times 10^{-7}$ s, $5.75 \times 10^{-4}$ s, $6.06 \times 10^{-4}$ s and $2.83 \times 10^{13}$ s. However, the desorption free energy for the interior layers of HAp is above 100 kcal/mol which leads to the required barrier crossing time of $2.6 \times 10^{54}$ s, indicating that it is practically impossible. This leaves the surface ions the only candidates for vacancy creation and ion doping at pH 5 and 7 without demineralization. Thus, the values of the activation free energy obtained through SMD simulations are in line with the finding of ion doping in the surface obtained through the conventional MD simulation.

$$t = \frac{1}{Ae^{-\frac{E_a}{kT}}} \tag{8}$$

Another important implication of this finding is that the ion doping in the interior of HAp can only occur during the mineralization process, in which the ions can be doped on the surface of the growing mineral.

Assuming that chemical stability is more important than the mechanical stability in the tooth enamel, it remained hitherto unclear which ions at what concentration can serve the goal of improving HAp chemical stability. To elucidate this aspect of ion doping, alchemical transformation and uniaxial compression simulations were deployed on HAp systems with various concentrations of several ions. It was found that F$^-$ doping for concentrations as high as 50% affect neither the mechanical stability nor the chemical stability. This is in line with the findings of Wang *et al.* [36] on inefficacy of F$^-$ in protecting the 001 facets and the lack of change in the mechanical properties due to F$^-$ doping. Here, it was theoretically established that, among Mg$^{2+}$, F$^-$ and CO$_3^{2-}$ ions, magnesium has the largest impact on both the mechanical and chemical properties, while fluorine has the least influence. Even though the overall system charge remained neutral through distribution of the extra charge into the remaining phosphate ions, the observed trend can be attributed to the change in the local arrangement of partial charges. Substitution of Ca$^{2+}$ with Mg$^{2+}$ leads to a change in the partial charges from +1.5e to +1.8e (+0.3e, where e = 1.602 × 10$^{-19}$ C) while there is a change of charge from -2.2e to -2.0e (+0.2e) and -0.9e to -1.0e (-0.1e) for the substitution of PO$_4^{3-}$ with CO$_3^{2-}$ and OH$^-$ with F$^-$, respectively. Thus, the efficacy of a given ion in the doping process is dependent on its partial charges inside the HAp structure which in turn is dependent on the covalent and ionic nature of the interaction, and itself is affected by its various chemical properties including electronegativity and polarizability.

Another unexplored aspect of the HAp doping covered in the current study was the effect of solvation. To isolate the effect of water molecules from HAp in the chemical stability of doped HAps, the contribution of HAp to the total free energy change was also reported in the current study. In the case of doping with Mg$^{2+}$, water molecules decrease the chemical stability, while for the CO$_3^{2-}$ an opposite trend is observed. Thus, it is expected that Mg$^{2+}$ ion doping decreases the HAp solubility while the CO$_3^{2-}$ has an opposite effect. The latter observation is in agreement with the finding of Mohammad *et al.* [37] on the effect of B-carbonate presence in HAp.

The author in the current study deployed various computational approaches for each utilizing a different strategy to ensure its validity which is explained in the supplementary information. The authors are still aware that the present study may have shortcomings. In the current study, the partial charges of the Mg$^{2+}$ and F$^-$ ions were estimated based on their polarization, while the CHARMM general forcefield parameters were adapted for the CO$_3^{2-}$ ion. A better approach would have been to recalibrate the forcefield parameters for these ions, but this lies outside the scope of the current study. While the SMD results presented here are qualitatively reliable despite not yielding precise free energy values, future studies will address this limitation through constant-pH MD simulations at lower pH and a more detailed investigation of dopant incorporation and its potential effects on hydroxyapatite morphology.

## Concluding remarks

In the present study, a combination of conventional MD simulations, Steered Molecular Dynamics (SMD), Thermodynamics Integration (TI) and uniaxial compression test approaches were deployed to investigate the kinetics, thermodynamics and mechanics of ion doping in the hydroxyapatite (HAp) as the main mineral of tooth enamel. The simulation results identified that the surface was the only feasible location for ion doping, while pre-formed HAp clusters leaving the doping during the crystal growth was the only option for interior doping of HAp. Besides, it was revealed that $Mg^{2+}$ is more effective in increasing the chemical stability of HAp while it decreases the mechanical properties at pH values of 5 and 7. $F^-$ and $CO_3^{2-}$ were the other ions investigated in the current study which have either minimal or negligible effect on the HAp behaviour. This study provides guidelines for designing new doped HAp crystals with mechanical and chemical properties tailored for the specific needs of either tooth enamel or bone.

## Methods

The methodology is explained in several subsections. In the first subsection, the details of the model and the forcefield parameters are presented. Then, the methods are detailed used in each simulation type presented in the current study. Each type of MD simulation employed in the current study is reviewed, including the conventional MD simulation, Steered Molecular Dynamics (SMD), Thermodynamic Integration (TI) and uniaxial compression test simulation. Finally, the experimental approaches deployed in the current study are explained.

**Computational model setup**

The MD simulation system shown in Figure 4 was utilized in the current study to investigate the change in the mechanical and chemical stability of HAp caused by ionic substitution. The HAp model taken from the Interface forcefield database [38] was solvated in TIP3p water models (Figure 4a). Several ionic substitutions were considered in which calcium, hydroxyl and phosphate ions were partially replaced with magnesium, fluorine and carbonate ions (Figure 4b). The system depicted in Figure 4a was deployed to study the chemical stability, while the mechanical stability was investigated through systems with various sizes to avoid the periodic box effects (Figure 4c). More details on the forcefield parameters and the partial charges of ions is provided in the supplementary information.

**Simulation techniques**

Various simulation techniques were employed in this study. First, conventional MD simulations were conducted to explore the feasibility of ion doping within the simulation timescales. Subsequently, three different aspects of ion substitution were investigated using simulations in three distinct categories. The first aspect focuses on the initial step of ionic substitution, which involves the dissolution of ions from HAp to create space for the incoming ions. The free energy associated with this step, corresponding to the activation free energy, dictates the rate of ionic substitution and was investigated through SMD simulations, shedding light on the kinetics of the process. Next, the change in the chemical stability of the system provides insights into the thermodynamics of the process. This includes a comparison of the substitution capabilities of various ions and their effectiveness in protecting against chemical agents such as acids, studied through TI. Finally, mechanical stability simulations elucidate the resistance of HAp to mechanical loading, assessed *via* uniaxial compression test modelling. Each of these methods is explained in detail below.

*Conventional MD simulations of ion substitution*

Several MD simulations with $Mg^{2+}$ or $F^-$ ions in the solutions were performed with conventional MD simulations to investigate the possibility of ionic doping in the time scales accessible to the conventional MD simulations. In each case the distribution of ions near the surface was investigated through a moving average approach. The simulations were done in NPT ensemble with temperature and pressure controlled to 310 K and 1 bar.

*Kinetics of the ion substitution*

SMD was employed to pull different components of pure HAp away from the surface, allowing the obtention of the free energy of their dissociation. This free energy represents the activation energy of ionic substitution, providing valuable insights into the kinetics of the process. First, the HAp facets of (001) and (010) at pH values of 5 and 7 and solvated in the 0.1 M NaCl solution were equilibrated for 50 ns simulations twice (or for 30 ns simulations three times) with different seed numbers to create the initial random configurations for the SMD simulations. Afterwards, for each one of the HAp components of $Ca^{2+}_{(I)}$, $Ca^{2+}_{(II)}$, $Ca^{2+}_{(surface)}$, , $PO_4^{3-}$ and $PO_4^{3-}_{(surface)}$, 50-150 SMD simulations were run from the configurations sampled in the 50 ns simulations in which that component is pulled away from the surface with the pulling rate of 1 Å/ns for either 15 or 25 Å and then they are pulled back to their initial positions in the reverse direction. More details on the free energy calculations from the SMD simulations and the processing of the results are available in the supplementary information.

*Chemical Stability*

TI was deployed to calculate the chemical stability which is defined as the free energy difference between the original HAp system with its doped counterpart. In TI the free energy difference between the state A (for instance $Ca^{2+}$ for $Ca^{2+}$ substitution with $Mg^{2+}$) and state B ($Mg^{2+}$) is obtained through equation 7 in which $\Delta G_{AB}$, $U$ and $\lambda$ represent the free energy difference, potential energy and the control parameter whose value changes from 0 to 1 in switching from state A to state B. A new command in LAMMPS was written to calculate the value of $\partial U/\partial \lambda$ for each intermediate state. To verify that the distance between windows is small enough the value of $\partial U/\partial \lambda$ for smaller distances between windows was calculated which showed no difference. More details are available in the Supplementary Information (SI) document.

$$\Delta G_{AB} = \int_0^1 \langle \frac{\partial U}{\partial \lambda} \rangle_\lambda \tag{7}$$

Three different pathways for thermodynamic integration were utilized in the current study corresponding to each of the ion substitution candidates. The first pathway was related to the substitution of calcium ions with magnesium in which the values of non-bonded interaction parameters and charges were changed from the initial value to the final one. 51 windows with the spacing of $\delta \lambda = 0.02$ were used in the stratification. Since the charge difference of $F^-$ with $OH^-$ is lower than $Mg^{2+}$ with $Ca^{2+}$, in the second pathway for $F^-$ ion substitution only 10 windows were utilized. Since in this case the hydrogen atom of the hydroxyl group does not have any Lennard-Jones interactions with the other atoms and the effect of bonds in the free energy calculations is negligible [39], the hydrogen atom and its bonded interaction remained intact during the alchemical transformation. For all these cases the thermodynamic integration control parameter was defined as $\lambda_i = (\varepsilon_i - \varepsilon_{initial})/(\varepsilon_{final} - \varepsilon_{initial})$ in which $\varepsilon_i$, $\varepsilon_{initial}$ and $\varepsilon_{final}$ represent the Lennard-Jones epsilon value for the stratification window $i$, $Ca^{2+}$ (O) and $Mg^{2+}$ ($F^-$), respectively.

The third pathway is related to the alchemical transformation from $PO_4^{3-}$ to $CO_3^{2-}$ for which the situation is slightly different than the previous cases, as in this case one oxygen atom in the phosphate turned into dummy atoms and some bond and angle interactions were turned off. Due to the convergence issue near the end points in atom annihilation [40], the soft-core potential was used for this transformation. The bonds and angles were linearly scaled between the corresponding values for phosphate and carbonate ions for the interactions present in both the structures. The extra bonds and angles of the phosphate molecule were interpolated between their original values and the final value of zero (the single topology approach).

*Mechanical Stability*

Mechanical stability is defined as the ability of a system to resist mechanical deformation. Uniaxial compression simulation was deployed to measure the change in the mechanical stability of HAp caused by ionic substitution (Figure 4d). To eliminate the system size effect for each case, various sizes of the system were considered to find the length after which the response is independent of the system size. The strain rate of $10^5$ $s^{-1}$ was adapted here.

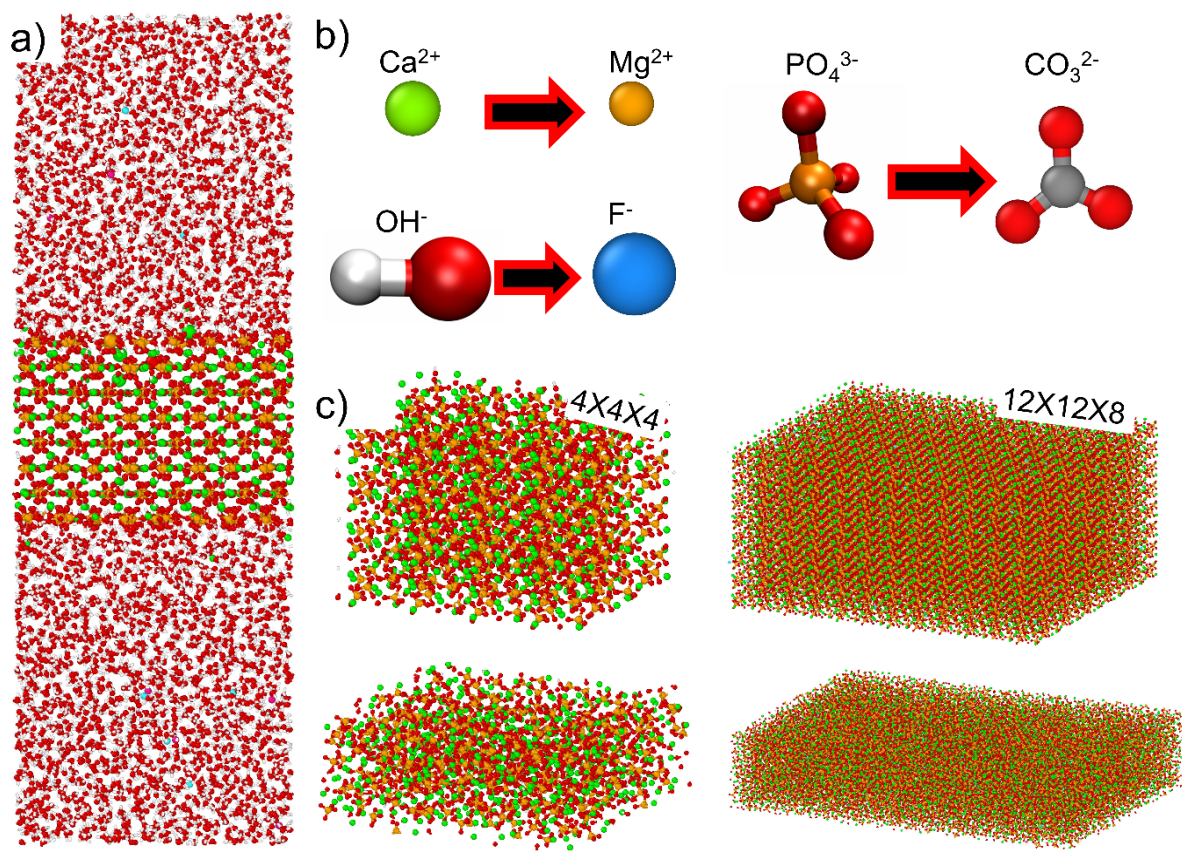

*Figure 4 – The MD simulation setup for the current study. a) The main system of HAp solvated in TIP3p water molecules with NaCl concentration of 0.1 M. Substitution of b) Calcium with Magnesium, Hydroxyl with Fluorine and Phosphate with Carbonate ions for which the chemical and mechanical stabilities are investigated. c) The simulation setup for studying the mechanical stability with computational boxes of different sizes. Due to the periodic boundary effect several different system sizes were used in studying the mechanical stability which are shown in the initial state (upper part of panel c) and the final compressed state (lower part of panel c).*

## Acknowledgements

The project made use of HPC time granted *via* the UK High-End Computing Consortium for Biomolecular Simulation, HECBioSim (http://hecbiosim.ac.uk), supported by EPSRC (grant no. EP/X035603/1). Via our membership of the UK's HEC Materials Chemistry Consortium (MCC) funded by EPSRC (EP/X035859), the present work used the ARCHER2 UK National Supercomputing Service (http://www.archer2.ac.uk). Some of the computations described in this paper were performed using the University of Birmingham's BlueBEAR HPC service, which provides a High-Performance Computing service to the University's research community. See http://www.birmingham.ac.uk/bear for more details.

## Author Contributions

M.T. designed the molecular dynamics simulation approach, selected and implemented relevant models, performed all simulations and data analysis, and drafted the manuscript. The computational methodology was developed beyond the outline provided in the original proposal. A.M.K. contributed input on experimental context and identified key parameters of interest. G.L. and R.M.S. provided feedback on manuscript drafts, mainly contributing to the introduction and general scientific framing and also provided access to the BlueBear HPC. J.C and C.B reviewed the manuscript. J.C.T. provided institutional support, facilitated access to UK HEC MCC and reviewed the final manuscript. A.M.K., G.L, R.M.S. and C.B. were involved in grant writing and funding acquisition.

All authors reviewed and approved the final version of the paper.

## Funding

AMK, JCT and all the co-authors wish to acknowledge the UK Engineering and Physical Sciences Research Council for the award of the principal funding for the present study under EPSRC grant no. EP/W009412/1 "Elucidating the pathways for human tooth enamel mineralization by 4D microscopy and microfluidics".

*Supplementary Information*

# Uncovering the role of ionic doping in hydroxyapatite: The building blocks of tooth enamel and bones

Mahdi Tavakol[1, 2], Jinke Chang[1], Cyril Besnard[1], Gabriel Landini[2], Richard M. Shelton[2], Jin-Chong Tan[1*], and Alexander M. Korsunsky[1,3**]

[1] Multi-Beam Laboratory for Engineering Microscopy (MBLEM) and Multifunctional Materials & Composites (MMC) Laboratories, Department of Engineering Science, University of Oxford OX1 3PJ, UK
[2] School of Dentistry, University of Birmingham, 5 Mill Pool Way, Edgbaston, Birmingham B5 7EG, UK
[3] Professor and Fellow Emeritus, Trinity College, Oxford OX1 3BH, UK

Corresponding authors:
[**]alexander.korsunsky@eng.ox.ac.uk
[*]jin-chong.tan@eng.ox.ac.uk

# Supplementary tables and figures

*Table S1 – The nonbonded interaction parameters used in the current study.*

| Atom | q | ε | σ | Ref |
|---|---|---|---|---|
| $Ca^{2+}$ | +1.5 | 0.1299984877 | 2.9399690301 | [1] |
| $O\ (PO_4^{3-})$ | -0.8 | 0.0699984928 | 3.0290604000 | |
| $P\ (PO_4^{3-})$ | 1.0 | 0.2800057201 | 3.8308580628 | |
| $O\ (OH^-)$ | -1.1 | 0.0800014955 | 3.2963200383 | |
| $H\ (OH^-)$ | 0.2 | 0.0000000000 | 0.0000000000 | |
| $PO_4^{3-}$ | -2.2 | ----------------- | ----------------- | |
| $OH^-$ | -0.9 | ----------------- | ----------------- | |
| $Mg^{2+}$ | 1.8 | 0.0150000000 | 2.1114299620 | [2] |
| $F^-$ | -1.0 | 0.0400000000 | 3.4121400000 | [3] |
| $C\ (CO_3^{2-})$ | 0.97 | 0.0175000000 | 3.5635948730 | [4] |
| $O\ (CO_3^{2-})$ | -0.99 | 0.0300000000 | 3.0290556420 | |
| $CO_3^{2-}$ | -2.0 | ----------------- | ----------------- | |

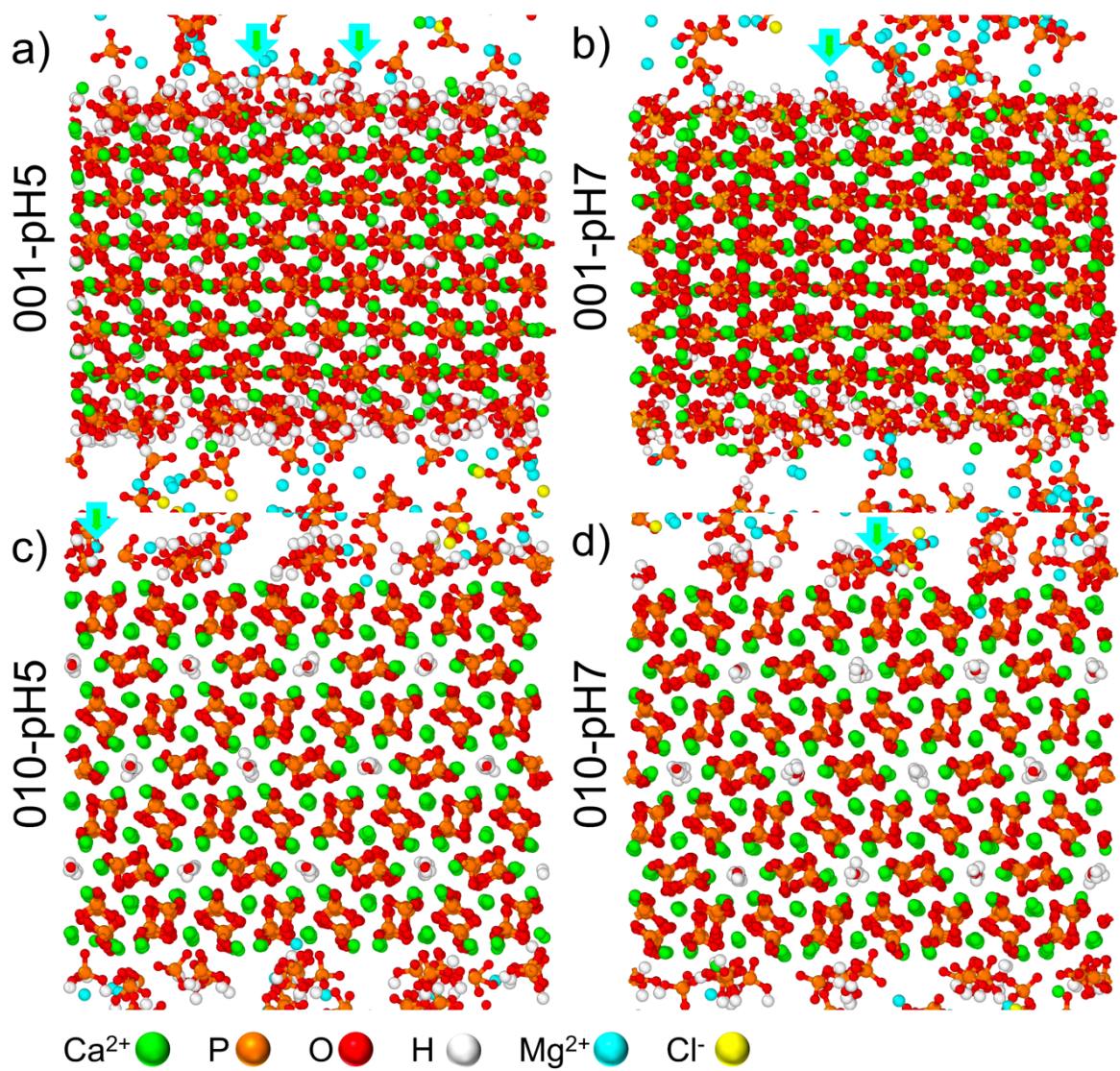

*Figure S1 – The results of the conventional MD simulation for the doping with $Mg^{2+}$ ions show doping in the surface of HAp for various facets and pH values. The final simulation snapshot for the a), b) 001, and c), d) 010 facets of HAp at pH of a), c) 5 and b), d) 7 solvated in 2 M solution of $Mg_3(PO_4)_2$. The doped $Mg^{2+}$ ions are indicated with arrows which all are located only on the surface.*

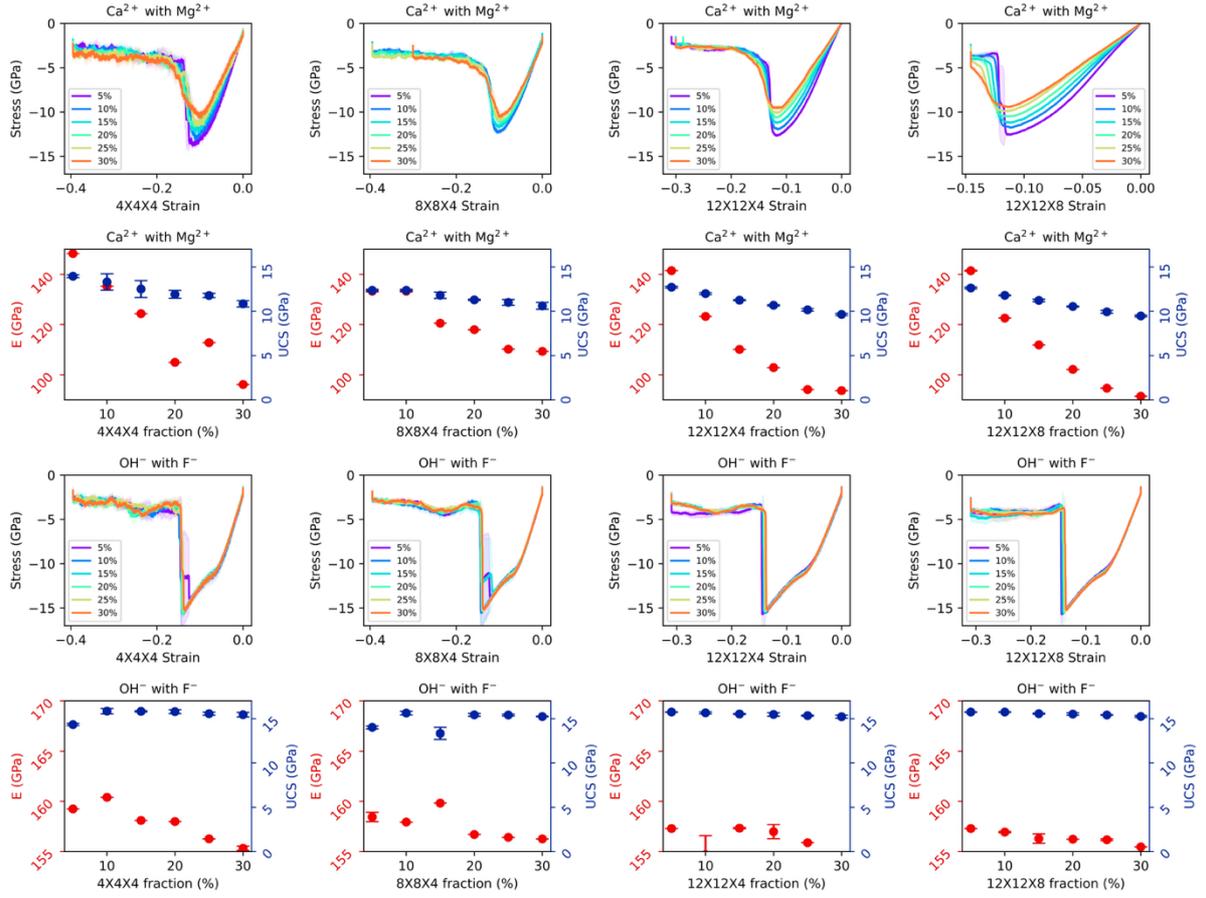

*Figure S2 – The stress-strain and mechanical properties plots for various ionic substitutions as a function of simulation box size showing convergence for boxes larger than 8×8×4 lattices.*

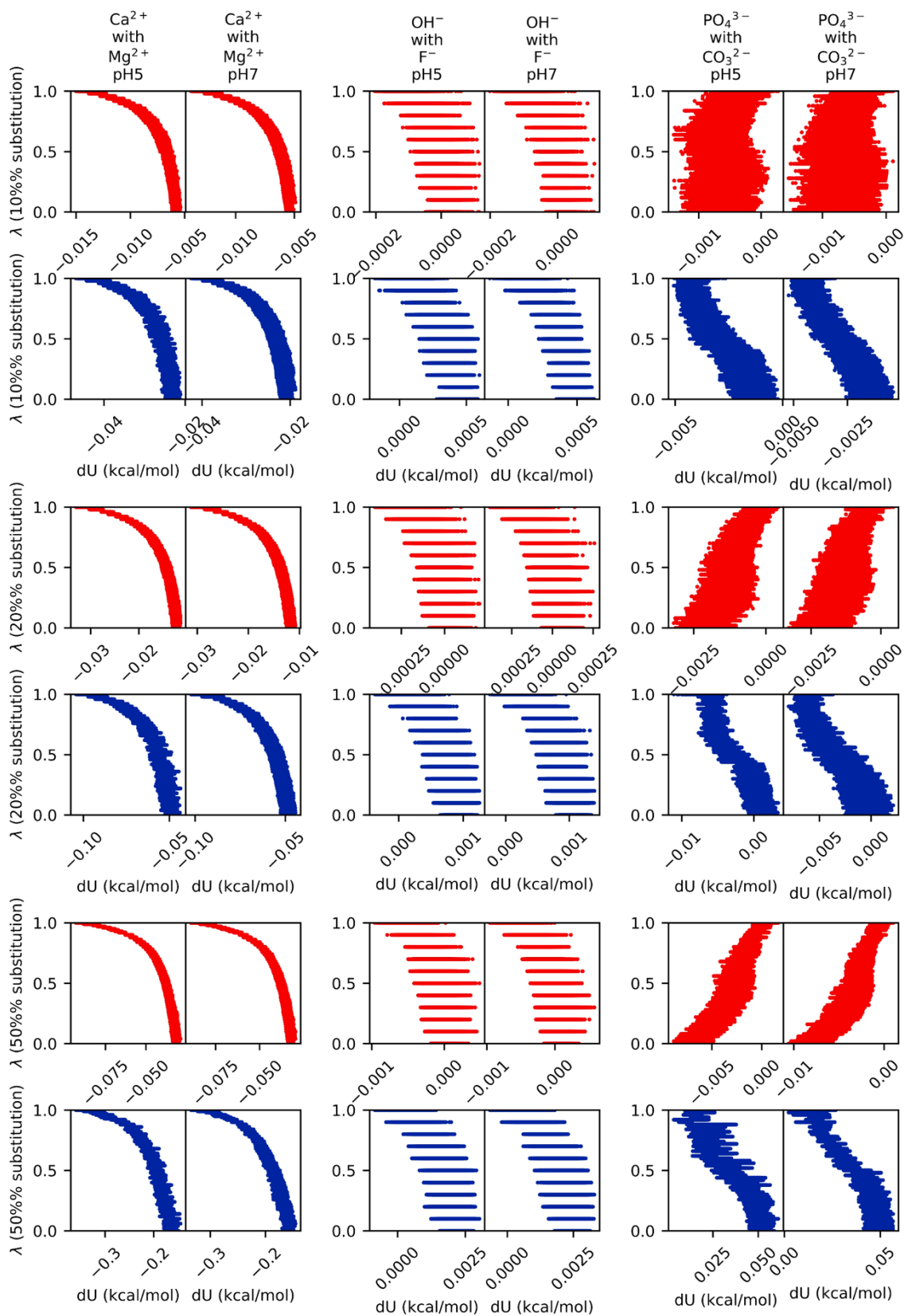

*Figure S3 – The distribution of dU for the thermodynamics integration for the whole system (red plots) and HAp (blue plots), showing enough overlap between neighboring windows validating the outcome of the thermodynamics integrations for various ion doping with different concentrations.*

**Supplementary information on the forcefield details**

A vital part of MD simulation is the interatomic forcefield used in calculating the interatomic interactions. In the current study, two different interatomic forcefields were employed. In all simulations except for the TI of $PO_4^{3-}$ to $CO_3^{2-}$ the Lennard-Jones (LJ) potential for non-bonded interactions and the Coulomb law for electrostatic interactions were utilized (equations 1, 2). However, when a site is created or annihilated in TI such as the alchemical transformation from $PO_4^{3-}$ to $CO_3^{2-}$, results diverge near the initial or the final structure [5]. Thus, soft-core alternatives for the LJ and Coulombic interactions were deployed in this case (equations 3, 4). Here, the parameters $U_{ij}, r_{ij}, \varepsilon, \sigma, C, q_i, \epsilon, \lambda, \alpha_{LJ}, \alpha_C$ and $n$ represent the potential energy, interatomic distance, LJ potential well, LJ diameter, Columb law constant, charge, dielectric constant, the alchemical transformation control parameter, LJ and Coulomb parameters for the soft-core potential and the soft-core exponent, respectively. In the current study the values of 2, 0.5 and 10 were chosen for the parameters $n$, $\alpha_{LJ}$ and $\alpha_C$, respectively.

$$U_{ij} = 4\varepsilon\left[\left(\frac{\sigma}{r_{ij}}\right)^{12} - \left(\frac{\sigma}{r_{ij}}\right)^6\right] \tag{1}$$

$$U_{ij} = \frac{C q_i q_j}{\epsilon r_{ij}} \tag{2}$$

$$U_{ij} = 4\varepsilon\lambda^n\left[\frac{1}{\left[\alpha_{LJ}(1-\lambda)^2 + \left(\frac{r_{ij}}{\sigma}\right)^6\right]^2} - \frac{1}{\alpha_{LJ}(1-\lambda)^2 + \left(\frac{r_{ij}}{\sigma}\right)^6}\right] \tag{3}$$

$$U_{ij} = \lambda^n \frac{C q_i q_j}{\epsilon\sqrt{\alpha_C(1-\lambda)^2 + r_{ij}^2}} \tag{4}$$

In the current study the interface forcefield parameters were deployed for the pure HAp [1]. Since the interaction between $Ca^{2+}$ with $PO_4^{3-}$ is not completely ionic and has a slight covalent character, the charge of the calcium ion is set to +1.5, slightly lower than its nominal charge of +2.0. Similarly, the charge of $OH^-$, is -0.9 which is slightly lower in magnitude than its nominal value of -1.0. In the current study, the parameter $q^2/r$ was used to judge the polarizability of ions and accordingly select their partial charges. With regards to the value of this parameter, the charge values of +1.8 and -1.0 were chosen for $Mg^{2+}$ and $F^-$ ions, respectively. The selection of non-bonded parameters is less challenging than the determination of partial charges. The non-bonded parameters for $Mg^{2+}$, $F^-$ and $CO_3^{2-}$ ions were adapted from the values reported in references [2-4]. The list of interatomic parameters for the ions in the current study is shown in Tabel S1.

**Supplementary information on free energy calculations through SMD simulations**

The pulling free energy can be calculated either through the Jarzynski equation (equation 5) or the Bennet Acceptance Ratio (BAR, equation 6) in which $\Delta G$, $W_{forward}$, $W_{reverse}$, $kT$, $\langle \ \rangle_0$ and $\langle \ \rangle_{0'}$ represent the free energy change, forward work, reverse work, the thermal energy, the ensemble average for the forward work and the ensemble average for the reverse work. Although, in principle, the ensemble average for the reverse process should be obtained from an equilibrium simulation in which the pulled atom is initially placed at its final position from the forward pulling, we omitted this equilibration step since our preliminary results indicated no significant differences.

The processing of the SMD simulation results was automated through some bash and C++ codes [6]. The *pymbar* module of python was utilized for the BAR analysis [7]. In the BAR formula the $W_{forward}$ is defined as the forward work moving from the initial SMD position of 0 to the final position of $z$, while the $W_{reverse}$ is the work in reverse direction from position $z$ to the position of 0 covering the same distance span but in the reverse direction.

$$e^{-\Delta G/kT} = \langle e^{-W_{forward}/kT}\rangle_0 \tag{5}$$

$$e^{-\Delta G/kT} = \frac{\left\langle\frac{1}{1 + e^{W_{forward}-\Delta G}}\right\rangle_0}{\left\langle\frac{1}{1 + e^{W_{reverse}+\Delta G}}\right\rangle_{0'}} \tag{6}$$

## Supplementary information for thermodynamic integrations

A new command was added to LAMMPS for thermodynamic integration. During equilibrating the system at $\lambda = \lambda_0$, the new command at every $n$ steps first backed up the atomic forces, non-bonded parameters and charges. Then, the non-bonded parameters and charges were changed to the value corresponding to $\lambda_0 - d\lambda$ and then the potential energy value was calculated ($u_A$). Then, the same procedure was repeated for $\lambda_0 + d\lambda$ to obtain $u_B$. Finally, the parameter $\frac{du}{d\lambda}(\lambda = \lambda_0) = \frac{u_B - u_A}{2d\lambda}$ was calculated. The alternative approach in the form of postprocessing is to first equilibrate the system at $\lambda = \lambda_0$ and then for each snapshot obtained from the trajectory change the value of charges and non-bonded parameters and calculate the $u_A$ and $u_B$ parameters to obtain the $\frac{dU}{d\lambda}$. Since in the current study the trajectories have output with single precision the alternative method has a single precision in contrary to the new method with the double precision.

To verify the new command the value of $u$ ($u_A(d\lambda = 0)$) from the command with the LAMMPS output from compute PI was compared. Figure S4 depicts values of the parameter $u$ for the whole system (Figure S4a and Figure S4c) and the HAp (Figure S4b and Figure S4d) for various ionic substitution alongside the values calculated from the LAMMPS output illustrating a good agreement (error of less than $10^{-4}$ %).

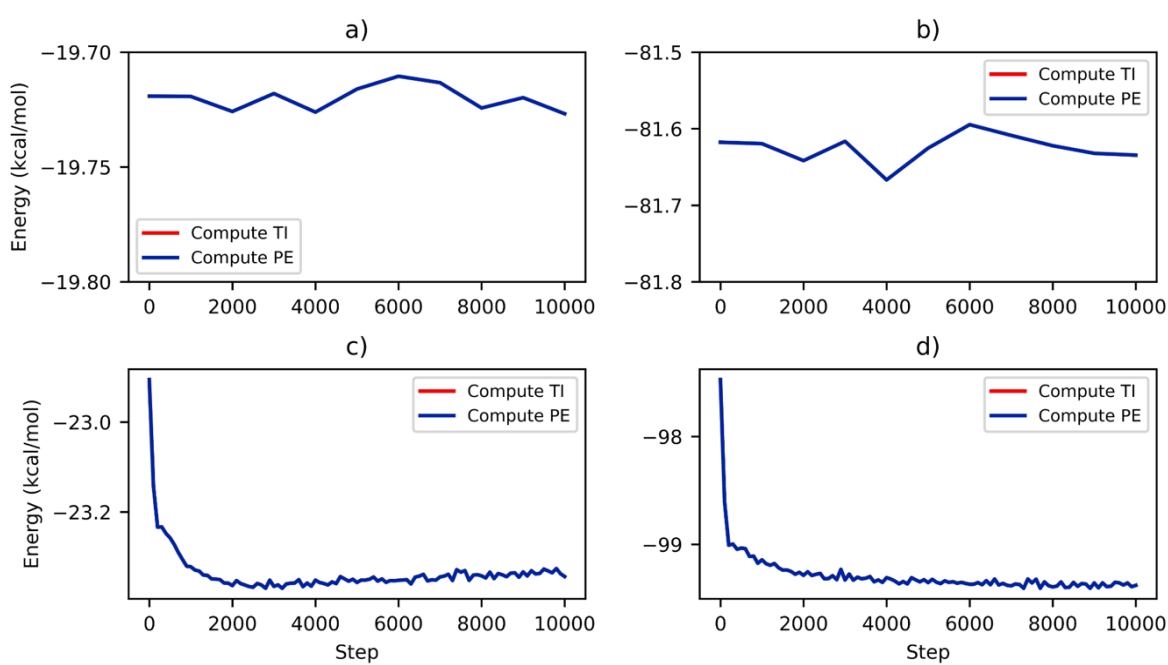

*Figure S4 - Comparison of the potential energy values obtained from the new command in the current study (compute TI) and the LAMMPS compute PE command for (a), (c) the whole system and (b), (d) HAp in the substitution of (a), (b) 10% of hydroxyl groups to fluoride and (c), (d) 50% of calcium ions to magnesium showing the agreement with the original method.*

To investigate the effect of numerical precision in the thermodynamics integration the results of the new command are compared with the outcome from the trajectory postprocessing with single precision. The values for substitution of hydroxyl with fluoride (Figure S5), and calcium with magnesium (Figure S6), show that even though the error in the values of $u_A$ and $u_B$ is less than 1 percent the error in the $\frac{dU}{d\lambda}$ can reach several 100 percent with a higher error for HAp. The reason for such a behavior is the differential is calculated through subtracting two almost identically large values from each other ($u_B - u_A$) and dividing the result by a very small value (($u_B - u_A)/d\lambda$) to obtain the $\frac{dU}{d\lambda}$. Thus, numerical precision has a strong effect on the result accuracy.

Finally, to verify the correctness of the calculations for $dU/d\lambda$, the result of the calculation with three values of $d\lambda = 0.02$, $d\lambda = 0.01$ and $d\lambda = 0.005$ are compared (
Figure S7). The outcome shows no effect of the $d\lambda$ parameter on the $dU/d\lambda$ parameter for the selected value of $d\lambda$ which illustrates that the chosen value of this parameter is small enough to get valid outcomes.

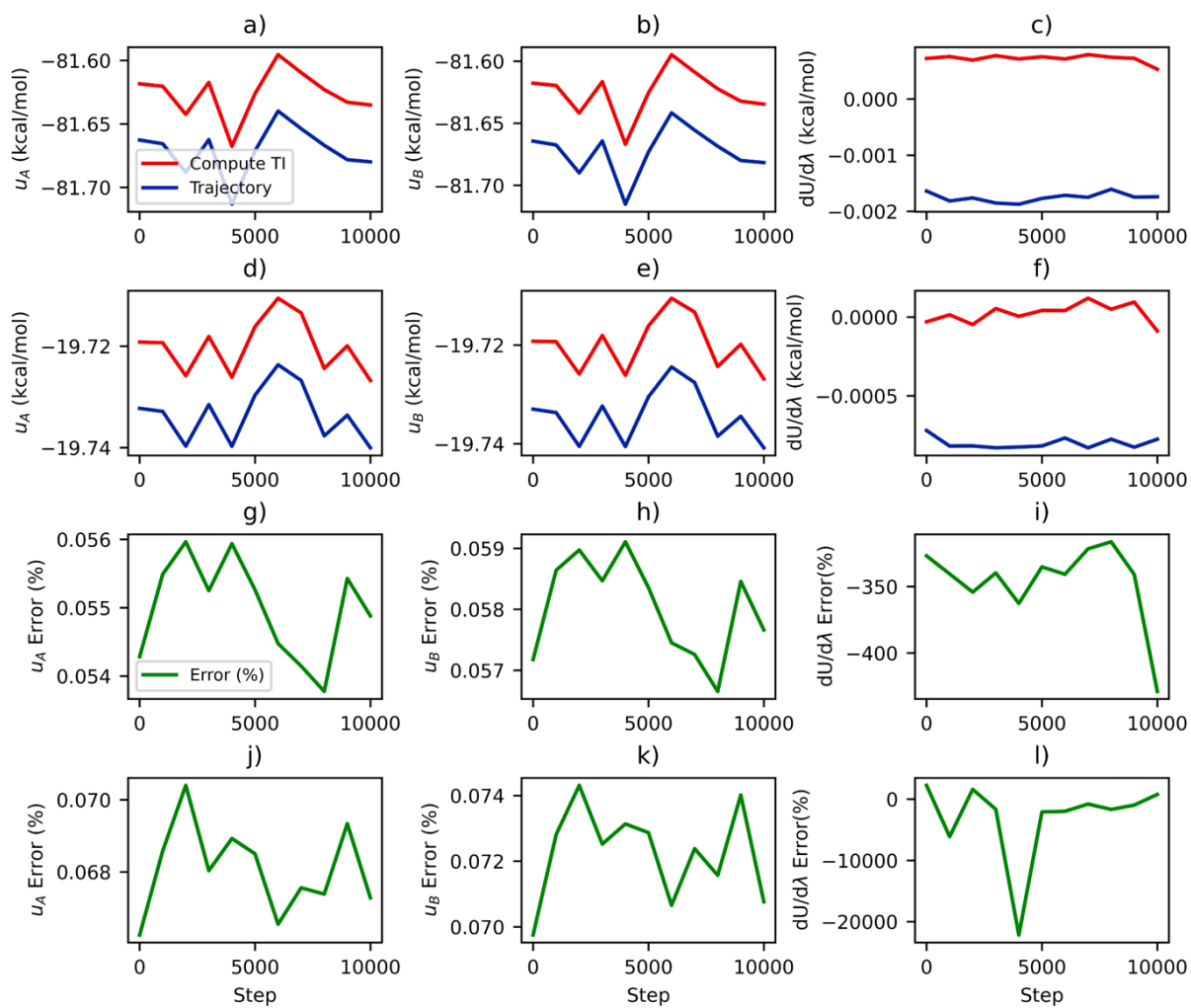

*Figure S5 – The numerical precision affects the thermodynamic integration outcomes. The values of (a), (d) $u_A$, (b), (e) $u_B$ (c), (f) $du/d\lambda$ and error values for (g), (h) $u_A$, (h), (?) $u_B$ (i), (l) $du/d\lambda$ for (a), (b), (c), (g), (h), (i) HAp and (d), (e), (f), (j), (k), (l) the whole system for substitution of 10% of hydroxyl groups with fluoride.*

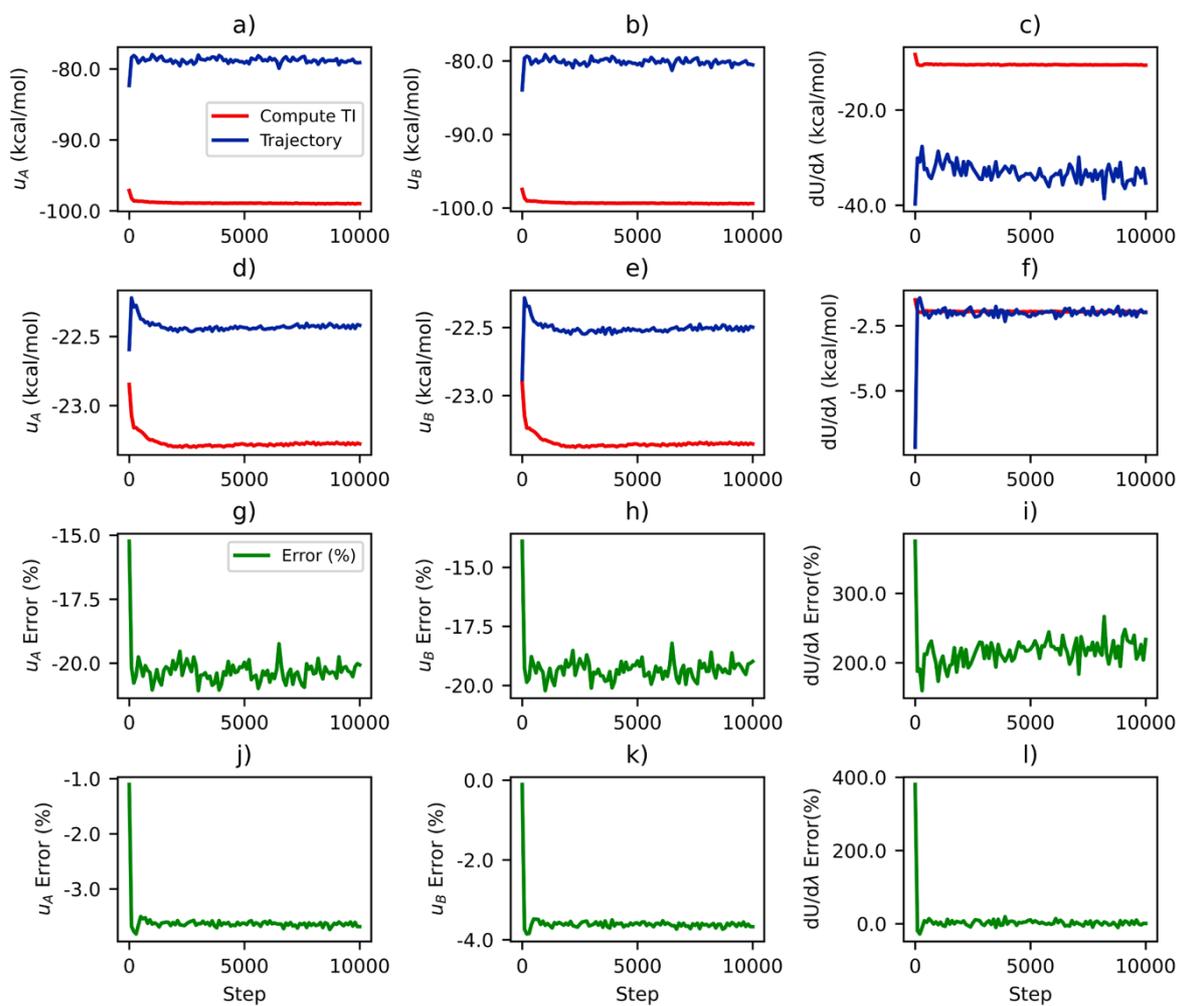

*Figure S6 - The numerical precision affects the thermodynamic integration outcomes. The values of (a), (d) $u_A$, (b), (e) $u_B$ (c), (f) $du/d\lambda$ and error values for (g), (h) $u_A$, (h), () $u_B$ (i), (l) $du/d\lambda$ for (a), (b), (c), (g), (h), (i) HAp and (d), (e), (f), (j), (k), (l) the whole system for substitution of 50% of calcium ions with magnesium.*

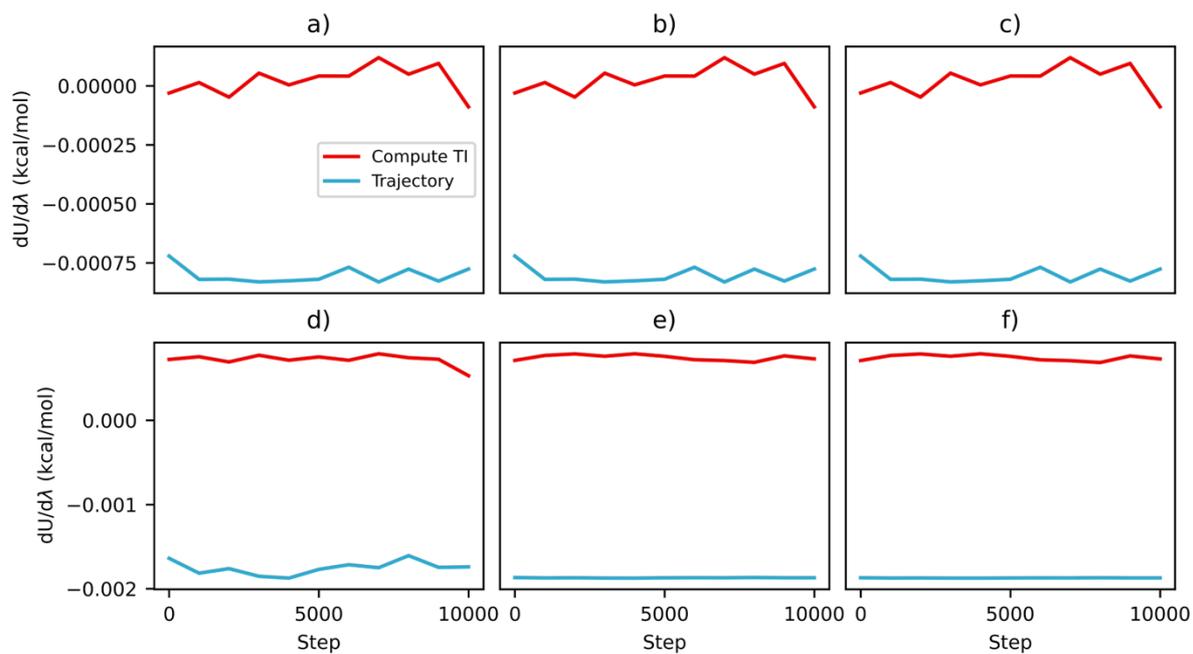

*Figure S7 – The values of dU/dλ for (a), (b), (c) HAp and (d), (e), (f) the whole system and various amounts of dλ of (a), (d) 0.02, (b), (d) 0.01 and (c), (f) 0.005.*

**Supplementary information for validation of results**

In the current study, computational approaches such as uniaxial compression, free energy calculations through SMD simulations, and TI were deployed, each requiring specific simulation strategy to ensure the validity of the results. To test the possibility of the artifacts caused by the periodic boundary in the uniaxial compression test, several box sizes were used which illustrated that for box sizes larger than a specific threshold, there is no size effect (Figure S2). All the values reported in the current study for mechanical stability are for the largest computational box. In addition to the statistical errors, the free energy calculations are also prone to sampling errors which are harder to identify since their roots are in the unvisited pathways or states [8]. For the SMD free energy calculations the results of BAR were compared with the average work obtained through 2 simulations with pulling speeds of 0.1 Å/ns and 0.01 Å/ns (10 and 100 times slower than the original one). The results for the 0.1 Å/ns and 0.01 Å/ns pulling simulations (ran for 150 days on a machine with a 48 core Intel Xeon Platinum 8268 CPU) were almost identical for pulling $Ca^{2+}_{(I)}$ from the 010 facet of HAp at pH 5 which shows that the reversible work regime was reached. Even though the BAR values for other cases can have errors as high as 31% with respect to the reversible work values (0.01 A /ns), this does not affect the conclusion regarding ion doping on the surface, since the difference between the activation free energy for surface and interior atoms is much higher than that. For the TI, sufficient overlap between neighboring windows is required to avoid biasing errors. In the current study the overlap was shown to be adequate from results shown in Figure S3.